# Quantitative scattering theory of near-field response for 1D polaritonic structures

**Lorenzo Orsini[1], Iacopo Torre[1], Hanan Herzig-Sheinfux[1] and Frank H. L. Koppens[1,2,*]**

1 ICFO-Institut de Ciencies Fotoniques; 08860 Castelldefels (Barcelona), Spain.
2 ICREA-Institució Catalana de Recerca i Estudis Avançats; 08010 Barcelona, Spain.
* frank.koppens@icfo.eu

## Abstract

Scattering-type scanning near-field optical microscopy is a powerful imaging technique for studying materials beyond the diffraction limit. However, interpreting near-field measurements poses challenges in mapping the response of polaritonic structures to meaningful physical properties. To address this, we propose a theory based on the transfer matrix method to simulate the near-field response of 1D polaritonic structures. Our approach provides a computationally efficient and accurate analytical theory, relating the near-field response to well-defined physical properties. This work enhances the understanding of near-field images and complex polaritonic phenomena. Finally, this scattering theory can extend to other systems like atoms or nanoparticles near a waveguide.

## 1. Introduction

Near-field optical microscopy is a powerful set of imaging techniques that allows for optically measuring and characterising objects with resolutions beyond the diffraction limit of traditional optical microscopes [1–6]. Specifically, the scattering-type scanning near-field optical microscope (s-SNOM) [5,6] has gained significant popularity over the last decades due to its exceptional spatial resolution, sensitivity to various optical phenomena, and its applicability in a wide range of the electromagnetic spectrum, from the visible to the THz. This technique provides insights into the optical properties of materials, enables nanoscale imaging, and finds applications in various fields, including chemistry [7,8], biology [9] and nanophotonics [10–12]. In nanophotonics, s-SNOM has played a critical role in unravelling the behaviour of polaritons in nanostructured surfaces [13–15]. Near-field measurements allow for the investigation of nanoscale optical properties and the characterisation of polaritonic systems such as nanocavities [16–18], polaritonic crystals [19,20] and other polaritonic structures [21–23]. However, interpreting the near-field data obtained through s-SNOM presents challenges due to the complexity of the measured observables. Mapping the near-field response of polaritonic structures while incorporating relevant physical properties remains a complex task. This challenge becomes even more significant when attempting to accurately interpret the gathered data quantitatively, which requires modelling the tip-substrate near-field interaction and demodulating the simulated signal accordingly.

A potential solution lies in numerical full-wave simulations incorporating realistic tip geometry. However, the significant scale difference between the wavelength, system size and tip-sample distance presents computational challenges, making tasks such as imaging with variable tip positions and spectroscopy with diverse light frequencies computationally demanding [24–27]. On the other hand, simplistic point-dipole models [5,28–30] do not adequately address the complexities inherent to the polaritonic systems we seek to understand. By addressing these challenges, we can achieve a more precise and accurate interpretation of complex phenomena occurring in our polaritonic systems, ultimately deepening our understanding of their underlying physics.

Here, we present a semi-analytical theory for the scattering near-field response of 1D polaritonic structures, bridging the need to find correspondence between experimental observables and the physical properties of those polaritonic systems. By extending the transfer matrix method to model the propagation and the near-field scattering process of polaritons in one-dimensional systems, we obtained a computationally efficient, physically meaningful, and yet accurate computation of the near-field observables. As a proof-of-concept demonstration, we investigate how the near-field response of periodic polaritonic systems corresponds to their band structure. Finally, we show that our scattering theory can calculate the near-field response of both pseudo-heterodyne and self-homodyne detection schemes, showcasing how the two schemes precisely relate to each other and how they are affected by unknown environmental variables of the measurement setup.

## 2. Formulation

The scattering theory presented in this work utilises the transfer matrix method (TMM) to model light propagation in one-dimensional systems. Within this framework, the theory employs the concept of the 3-port device to solve the complex near-field scattering problem. In general, those one-dimensional systems can be as generic as a slab or channel waveguides characterised by spatial modulation of their dielectric properties.

### 2.1 The polaritonic propagation problem

The TMM is widely used to solve the propagation of plane waves in one-dimensional multi-layered structures. In the case of electromagnetic waves, this method solves the one-dimensional Helmholtz equation with standard boundary conditions for electromagnetic waves. The method involves representing each layer as a matrix that describes the transmission and reflection of waves across that specific layer. By multiplying these matrices together, one can obtain the overall scattering matrix that defines the behaviour of the wave as it passes through all of the layers.

In our specific scenario, we model our polaritonic system as a layered structure by assigning an effective refractive index to each section. Fig. 1A illustrates this concept, where the top portion of the panel displays the calculated electric field profiles of a polariton propagating through a slab waveguide under various environmental conditions. As a result, the polariton exhibits different wavelengths in different sections of the structure, which can be correlated to an effective refractive index for each corresponding section (as shown in the bottom part of Figure 1A). We can formally write the effective refractive index of each section as follows:

$$n_{eff}(\omega) = \frac{c}{\omega \lambda_P(\omega)} \qquad 1$$

Here, $n_{eff}$ represents the effective refractive index of the specific section, $\omega$ denotes the light frequency, $c$ stands for the speed of light, and $\lambda_P$ is the polariton wavelength evaluated at frequency $\omega$. In practice, for each layer in the structure, we can determine a solution for the propagating polaritonic electric field along the structure dimension $x$, which can be expressed as:

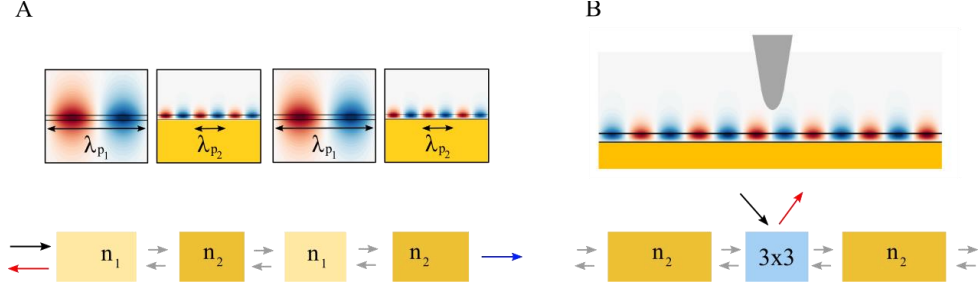

**Fig. 1, Schematic representation of the standard TMM and the 3x3 matrix variant.**
(**A**) The top part of the panel illustrates polaritons in two distinct environmental conditions, showcasing the impact of these conditions on the polaritonic wavelengths. This is depicted in the schematic shown in the bottom part of the panel. The schematic represents a block diagram of the structure, modelling the polaritonic system as a layered system with each section characterised by specific effective refractive indices, $n_1$ and $n_2$. According to the standard usage of TMM, the system is probed from the left side of the structure (indicated by a black arrow), and the scattered reflected (red arrow) and transmitted (blue arrow) amplitudes are calculated. (**B**) The top part of the panel displays a scenario where the near-field probe is in close proximity to the polaritonic media, allowing the tip apex to interact with the evanescently decaying tails of the polaritons that propagate through the medium. The bottom part of the panel presents a block diagram representing the tip-substrate system as a whole; specifically, the region of the tip apex is defined as a 3x3 scattering matrix. In terms of the TMM, the system is probed from the "far-field side" of the 3x3 matrix (indicated by a black arrow), and the near-field observable can be extracted by analysing the reflection amplitude (represented by a red arrow).

$$E_n(x) = E_n^+ e^{ik_n x} + E_n^- e^{-ik_n x} \qquad 2$$

Here, $n$ is the layer index, $E_n^\pm$ are the propagating (+) and counter-propagating (-) field amplitudes, and $k_n$ is the polariton propagation wavenumber, which corresponds to $k_n = 2\pi/\lambda_P$. The standard TMM precisely maps the forward- and back-propagating field amplitudes of the $n^{th}$ layer $(E_{n+1}^+, E_{n+1}^-)$ to the forward- and back-propagating field amplitudes of the $(n+1)^{th}$ layer $(E_n^-, E_{n+1}^+)$ via a 2x2 scattering matrix $S_n$ defined as follow:

$$S_n = \begin{bmatrix} S_{11}^n & S_{12}^n \\ S_{21}^n & S_{22}^n \end{bmatrix} \qquad 3$$

$$\begin{bmatrix} S_{11}^n & S_{12}^n \\ S_{21}^n & S_{22}^n \end{bmatrix} \begin{bmatrix} E_n^+ \\ E_{n+1}^- \end{bmatrix} = \begin{bmatrix} E_n^- \\ E_{n+1}^+ \end{bmatrix} \qquad 4$$

Then, to calculate the response of the entire system composed of an arbitrary number of N of layers, the scattering matrices $S_n$ are multiplied together using the Redheffer matrix product ($\otimes$).

$$S_G = S_1 \otimes \ldots \otimes S_N \qquad 5$$

Here, $S_G$ represents the global scattering matrix, which encompasses all the pertinent information regarding the system's response to incoming incident fields from the left or right boundaries of the structure. With the global scattering matrix, it is possible to calculate the transmission and reflection coefficients, which are ultimately connected to significant properties of the polaritonic system under study.

## 2.2 The near-field scattering problem

### 2.2.1 The s-SNOM

Scattering type scanning near-field optical microscopy (s-SNOM) is a type of scanning probe microscopy that enables subwavelength optical imaging with nanometre-scale resolution. To achieve this spatial resolution, a metalised atomic force microscope (AFM) tip undergoes sub-wavelength harmonic oscillations above the substrate. In many cases, an effective dipole model describes this system very well [5].

$$\vec{p}_{eff} = \alpha_{eff} \cdot \vec{E}_i \qquad 6$$

Here, $\vec{p}_{eff}$ is the amplitude of such an effective dipole. This quantity is proportional to the amplitude of the impinging electromagnetic wave $\vec{E}_i$ and to the polarizability of the effective tip-substrate system, denoted as $\alpha_{eff}$. Thus, the effective polarizability captures all the physics of the near-field interaction. Furthermore, the effective dipole emission conveys information about the relation between the local optical properties of the sample and the far-field, and it can be expressed as follows:

$$\vec{E}_s \propto \vec{p}_{eff} \qquad 7$$

Here, $\vec{E}_s$ represents the amplitude of the scattered electric field that reaches the s-SNOM photodetector. The field $\vec{E}_s$ depends on various factors, including but not limited to the distance between the photodetector and the tip, the incident polarisation, and the optical losses along the path. However, in an s-SNOM setup, these quantities are often constant, allowing us to simplify the dipolar emission in s-SNOM as follows:

$$E_s = \alpha_{eff} E_i \qquad 8$$

Here, $E_s$ and $E_i$ represent the scalar amplitudes of the scattered and incident fields, respectively, at the detector and the tip apex. Of particular significance, $\alpha_{eff}$ is a scalar that encompasses the effective polarizability of the dipolar system formed by the tip and the substrate, as well as the resulting dipolar emission in the far-field.

### 2.2.2 Integration in the TMM formalism

In general, we can conceptualise the process described in equation (8) for a polaritonic system by breaking down the near-field interaction into three steps:

1. Light couples into the structure at the tip location.

2. The polaritons propagate back and forth through the structure.

3. The polaritons couple out from the structure at the tip location.

This entire process can be intuitively modelled within the TMM framework by introducing a scattering element with an additional transmission/reflection channel. This approach is commonly used in microwave electronics [31], where complex devices with multiple input/output ports are modelled by $NxN$ scattering matrices, where $N$ denotes the number of ports. In our case, we model the near-field probe as a $3x3$ scattering matrix to account for the extra transmission/reflection channel. As shown in Fig. 1B, the polaritons can interact with the tip apex, either being reflected or transmitted in the polaritonic waveguide (1st and 2nd ports), or they can scatter into the far-field (3rd port).

In the following, we extend the standard TMM formulation to incorporate this scattering matrix and simulate the scanning process by computing the far-field reflection amplitude and phase of our polaritonic system at any given location of the near-field probe.

## 2.3 Formulation of the 3-port TMM

In the following, we implement a 3-port-device ($3x3$ scattering matrix) into a system composed of 2x2 scattering matrices. We start by defining that the top, left, and right sides of the 3-port-device are connected to three 2x2 scattering matrices that represent, respectively, the far-field channel (9) and the polaritonic left (10) and right (11) channels:

$$\begin{bmatrix} T_{11} & T_{12} \\ T_{21} & T_{22} \end{bmatrix} \begin{bmatrix} \alpha_T \\ A_T \end{bmatrix} = \begin{bmatrix} \beta_T \\ B_T \end{bmatrix} \qquad 9$$

$$\begin{bmatrix} L_{11} & L_{12} \\ L_{21} & L_{22} \end{bmatrix} \begin{bmatrix} A_L \\ \alpha_L \end{bmatrix} = \begin{bmatrix} B_L \\ \beta_L \end{bmatrix} \qquad 10$$

$$\begin{bmatrix} R_{11} & R_{12} \\ R_{21} & R_{22} \end{bmatrix} \begin{bmatrix} \alpha_R \\ A_R \end{bmatrix} = \begin{bmatrix} \beta_R \\ B_R \end{bmatrix} \qquad 11$$

Here we make a conceptual distinction between the field amplitudes written with capital Latin letters $A_R, A_L, A_T, B_R, B_L, B_T$ and the lower-case Greek letters $\alpha_R, \alpha_L, \alpha_T, \beta_R, \beta_L, \beta_T$. The first group represents the incident ($A_{R,L,T}$) and scattered ($B_{R,L,T}$) electric field amplitudes from the outer boundaries of the system, whereas the second group represents the incident ($\beta_{R,L,T}$) and scattered ($\alpha_{R,L,T}$) electric field amplitudes at the 3-port-device:

$$\begin{bmatrix} S_{11} & S_{12} & S_{13} \\ S_{21} & S_{22} & S_{23} \\ S_{31} & S_{32} & S_{33} \end{bmatrix} \begin{bmatrix} \beta_T \\ \beta_L \\ \beta_R \end{bmatrix} = \begin{bmatrix} \alpha_T \\ \alpha_L \\ \alpha_R \end{bmatrix} \qquad 12$$

The amplitudes written with Greek letters are the unknown-intermediate amplitudes that we want to get rid of. To do so, we must combine the previous equations (9, 10, 11) in the following matrix form:

$$\begin{bmatrix} L_{11} & L_{12} & 0 & 0 & 0 & 0 \\ L_{21} & L_{22} & 0 & 0 & 0 & 0 \\ 0 & 0 & T_{11} & T_{12} & 0 & 0 \\ 0 & 0 & T_{21} & T_{22} & 0 & 0 \\ 0 & 0 & 0 & 0 & R_{11} & R_{12} \\ 0 & 0 & 0 & 0 & R_{21} & R_{22} \end{bmatrix} \begin{bmatrix} A_L \\ \alpha_L \\ \alpha_T \\ A_T \\ \alpha_R \\ A_R \end{bmatrix} = \begin{bmatrix} B_L \\ \beta_L \\ \beta_T \\ B_T \\ \beta_R \\ B_R \end{bmatrix}$$

After rearranging the terms by interchanging rows and columns, we obtain the following:

$$\begin{bmatrix} T_{22} & 0 & 0 & T_{21} & 0 & 0 \\ 0 & L_{11} & 0 & 0 & L_{12} & 0 \\ 0 & 0 & R_{22} & 0 & 0 & R_{21} \\ T_{12} & 0 & 0 & T_{11} & 0 & 0 \\ 0 & L_{21} & 0 & 0 & L_{22} & 0 \\ 0 & 0 & R_{12} & 0 & 0 & R_{11} \end{bmatrix} \begin{bmatrix} A_T \\ A_L \\ A_R \\ \alpha_T \\ \alpha_L \\ \alpha_R \end{bmatrix} = \begin{bmatrix} B_T \\ B_L \\ B_R \\ \beta_T \\ \beta_L \\ \beta_R \end{bmatrix}$$

Then after substituting (12), we get:

$$\begin{bmatrix} T_{22} & 0 & 0 & T_{21} & 0 & 0 \\ 0 & L_{11} & 0 & 0 & L_{12} & 0 \\ 0 & 0 & R_{22} & 0 & 0 & R_{21} \\ T_{12} & 0 & 0 & T_{11} & 0 & 0 \\ 0 & L_{21} & 0 & 0 & L_{22} & 0 \\ 0 & 0 & R_{12} & 0 & 0 & R_{11} \end{bmatrix} \begin{bmatrix} 1 & 0 & 0 & 0 & 0 & 0 \\ 0 & 1 & 0 & 0 & 0 & 0 \\ 0 & 0 & 1 & 0 & 0 & 0 \\ 0 & 0 & 0 & S_{11} & S_{12} & S_{13} \\ 0 & 0 & 0 & S_{21} & S_{22} & S_{23} \\ 0 & 0 & 0 & S_{31} & S_{32} & S_{22} \end{bmatrix} \begin{bmatrix} A_T \\ A_L \\ A_R \\ \beta_T \\ \beta_L \\ \beta_R \end{bmatrix} = \begin{bmatrix} B_T \\ B_L \\ B_R \\ \beta_T \\ \beta_L \\ \beta_R \end{bmatrix} \quad 13$$

Now, we simplify this matrix equation by using the following notation:

$$S_{3PD} = \begin{bmatrix} S_{11} & S_{12} & S_{13} \\ S_{21} & S_{22} & S_{23} \\ S_{31} & S_{32} & S_{33} \end{bmatrix} \quad 14$$

$$\bar{\bar{M}}_{11} = \begin{bmatrix} T_{22} & 0 & 0 \\ 0 & L_{11} & 0 \\ 0 & 0 & R_{22} \end{bmatrix} \quad 15$$

$$\bar{\bar{M}}_{12} = \begin{bmatrix} T_{21} & 0 & 0 \\ 0 & L_{12} & 0 \\ 0 & 0 & R_{21} \end{bmatrix} \quad 16$$

$$\bar{\bar{M}}_{21} = \begin{bmatrix} T_{12} & 0 & 0 \\ 0 & L_{21} & 0 \\ 0 & 0 & R_{12} \end{bmatrix} \quad 17$$

$$\bar{\bar{M}}_{22} = \begin{bmatrix} T_{11} & 0 & 0 \\ 0 & L_{22} & 0 \\ 0 & 0 & R_{11} \end{bmatrix} \quad 18$$

$$\bar{A} = \begin{bmatrix} A_T \\ A_L \\ A_R \end{bmatrix}, \quad \bar{B} = \begin{bmatrix} B_T \\ B_L \\ B_R \end{bmatrix}, \quad \bar{\alpha} = \begin{bmatrix} \alpha_T \\ \alpha_L \\ \alpha_R \end{bmatrix}, \quad \bar{\beta} = \begin{bmatrix} \beta_T \\ \beta_L \\ \beta_R \end{bmatrix}, \quad 19$$

Therefore, equation (13) becomes more compact:

$$\begin{bmatrix} \bar{\bar{M}}_{11} & \bar{\bar{M}}_{12} \\ \bar{\bar{M}}_{21} & \bar{\bar{M}}_{22} \end{bmatrix} \begin{bmatrix} \mathbb{1}_{3\times 3} & \mathbb{0}_{3\times 3} \\ \mathbb{0}_{3\times 3} & S_{3PD} \end{bmatrix} \begin{bmatrix} \bar{A} \\ \bar{\beta} \end{bmatrix} = \begin{bmatrix} \bar{B} \\ \bar{\beta} \end{bmatrix} \quad 20$$

Now we need to solve the system of matrices in (20) to obtain an equation that looks like the following:

$$\bar{\bar{S}}_G \cdot \bar{A} = \bar{B} \quad 21$$

That is because $\bar{A}$ and $\bar{B}$ are the vectors of the incident and scattered field amplitudes, respectively, and, by definition, the global scattering matrix must map $\bar{A}$ into $\bar{B}$.

We begin by solving the matrix equation (20):

$$\begin{bmatrix} \bar{\bar{M}}_{11} & \bar{\bar{M}}_{12} \cdot S_{3PD} \\ \bar{\bar{M}}_{21} & \bar{\bar{M}}_{22} \cdot S_{3PD} \end{bmatrix} \begin{bmatrix} \bar{A} \\ \bar{\beta} \end{bmatrix} = \begin{bmatrix} \bar{B} \\ \bar{\beta} \end{bmatrix}$$

$$\begin{bmatrix} \bar{\bar{M}}_{11} & \bar{\bar{M}}_{12} \cdot S_{3PD} \\ \bar{\bar{M}}_{21} & \bar{\bar{M}}_{22} \cdot S_{3PD} - \mathbb{1}_{3\times 3} \end{bmatrix} \begin{bmatrix} \bar{A} \\ \bar{\beta} \end{bmatrix} = \begin{bmatrix} \bar{B} \\ \bar{0} \end{bmatrix}$$

Therefore:

$$\begin{cases} \bar{\bar{M}}_{11}\bar{A} + (\bar{\bar{M}}_{12} \cdot S_{3PD})\bar{\beta} = \bar{B} \\ \bar{\bar{M}}_{21}\bar{A} + [(\bar{\bar{M}}_{22} \cdot S_{3PD}) - \mathbb{1}_{3\times3}]\bar{\beta} = \bar{0} \end{cases}$$

Let us consider the second equation of that system:

$$[\mathbb{1}_{3\times3} - (\bar{\bar{M}}_{22} \cdot S_{3PD})]\bar{\beta} = \bar{\bar{M}}_{21}\bar{A}$$

$$\bar{\beta} = [\mathbb{1}_{3\times3} - (\bar{\bar{M}}_{22} \cdot S_{3PD})]^{-1} \cdot \bar{\bar{M}}_{21}\bar{A}$$

Now we can substitute $\bar{\beta}$ back to the first equation of the system and obtain:

$$\bar{\bar{M}}_{11}\bar{A} + (\bar{\bar{M}}_{12} \cdot S_{3PD}) \cdot [\mathbb{1}_{3\times3} - (\bar{\bar{M}}_{22} \cdot S_{3PD})]^{-1} \cdot \bar{\bar{M}}_{21}\bar{A} = \bar{B} \qquad 22$$

Therefore:

$$\bar{\bar{S}}_G = \bar{\bar{M}}_{11} + (\bar{\bar{M}}_{12} S_{3PD})[\mathbb{1}_{3\times3} - (\bar{\bar{M}}_{22} \cdot S_{3PD})]^{-1}\bar{\bar{M}}_{21} \qquad 23$$

Equation (23) solves the scattering problem of a generic combination of input amplitudes collected in $\bar{A}$. To complete the derivation, we can rewrite equation (21) in its explicit form and get the following:

$$\begin{bmatrix} S_{11}^G & S_{12}^G & S_{13}^G \\ S_{21}^G & S_{22}^G & S_{23}^G \\ S_{31}^G & S_{32}^G & S_{33}^G \end{bmatrix} \begin{bmatrix} A_T \\ A_L \\ A_R \end{bmatrix} = \begin{bmatrix} B_T \\ B_L \\ B_R \end{bmatrix} \qquad 24$$

Here, if we consider a situation in which the incident light is coming only from the far-field input of the 3-port-device, we get:

$$B_T = S_{11}^G \cdot A_T \qquad 25$$

This equation represents the main result of the implementation of a 3-port device in a standard TMM, and we conceptually recover the solution of the near-field scattering problem that we seek to solve. We find that equation (25) is equivalent to equation (8) because:

$$S_{11}^G = \alpha_{eff}, \qquad B_T = E_s, \qquad A_T = E_i \qquad 26$$

## 3. Scattering matrix of a 3-port-device

Now that we have successfully implemented the 3-port device into the standard formulation of the TMM, we formalise how a generic 3x3 matrix can be modelled as a scattering element, in our case, a near-field probe.

### 3.1 Formulation

A generic 3-port-device scattering matrix is written as follows:

$$S_{3PD} = \begin{bmatrix} S_{11} & S_{12} & S_{13} \\ S_{21} & S_{22} & S_{23} \\ S_{31} & S_{32} & S_{33} \end{bmatrix} \qquad 27$$

This object maps the incident electric fields of each port $(E_T^i, E_L^i, E_R^i)$ to the scattered ones $(E_T^s, E_L^s, E_R^s)$ as follows:

$$\begin{bmatrix} S_{11} & S_{12} & S_{13} \\ S_{21} & S_{22} & S_{23} \\ S_{31} & S_{32} & S_{33} \end{bmatrix} \begin{bmatrix} E_T^i \\ E_L^i \\ E_R^i \end{bmatrix} = \begin{bmatrix} E_T^s \\ E_L^s \\ E_R^s \end{bmatrix} \qquad 28$$

At this stage, the 3-port device can represent any scattering object that interacts with light propagating across a one-dimensional structure. For example, it can model an atom positioned on a waveguide, a nanoparticle coupled to a 1D polaritonic system, or the s-SNOM tip in the vicinity of a polaritonic medium.

Before modelling the behaviour of the specific object we aim to describe, we can apply some physical symmetries to reduce the degrees of freedom of the model. Actually, this complex-valued matrix (27) has a total of 18 degrees of freedom. In our specific case, we seek to model an elastic scattering process of light impinging on the s-SNOM tip apex. Therefore, we must enforce three important physical symmetries:

1. Mirror symmetry – the tip is assumed to be symmetric.
2. Energy conservation – the scattering process conserves energy.
3. Time reversal symmetry – the scattering process is a time reversal process.

### 3.1.1 Mirror symmetry

Let's first apply the mirror symmetry. A general matrix is mirror symmetric when the following equation is satisfied:

$$S_{3PD} = M^{-1} S_{3PD} M \qquad 29$$

Here, M is the mirror transformation matrix:

$$M = M^{-1} = \begin{bmatrix} 1 & 0 & 0 \\ 0 & 0 & 1 \\ 0 & 1 & 0 \end{bmatrix} \qquad 30$$

Therefore:

$$M^{-1} S_{3PD} M = \begin{bmatrix} 1 & 0 & 0 \\ 0 & 0 & 1 \\ 0 & 1 & 0 \end{bmatrix} \begin{bmatrix} S_{11} & S_{12} & S_{13} \\ S_{21} & S_{22} & S_{23} \\ S_{31} & S_{32} & S_{33} \end{bmatrix} \begin{bmatrix} 1 & 0 & 0 \\ 0 & 0 & 1 \\ 0 & 1 & 0 \end{bmatrix}$$

$$= \begin{bmatrix} 1 & 0 & 0 \\ 0 & 0 & 1 \\ 0 & 1 & 0 \end{bmatrix} \begin{bmatrix} S_{11} & S_{13} & S_{12} \\ S_{21} & S_{23} & S_{22} \\ S_{31} & S_{33} & S_{32} \end{bmatrix} = \begin{bmatrix} S_{11} & S_{13} & S_{12} \\ S_{31} & S_{33} & S_{32} \\ S_{21} & S_{23} & S_{22} \end{bmatrix}$$

By explicitly plugging this result in equation (29), we obtain:

$$\begin{bmatrix} S_{11} & S_{12} & S_{13} \\ S_{21} & S_{22} & S_{23} \\ S_{31} & S_{32} & S_{33} \end{bmatrix} = \begin{bmatrix} S_{11} & S_{13} & S_{12} \\ S_{31} & S_{33} & S_{32} \\ S_{21} & S_{23} & S_{22} \end{bmatrix}$$

Therefore:

1. $S_{11} \stackrel{\text{def}}{=} b e^{i\varphi_b}$ $\in \mathbb{C}$
2. $S_{31} = S_{21} \stackrel{\text{def}}{=} c_i e^{i\varphi_i}$ $\in \mathbb{C}$
3. $S_{13} = S_{12} \stackrel{\text{def}}{=} c_o e^{i\varphi_o}$ $\in \mathbb{C}$

4. $S_{33} = S_{22} \stackrel{\text{def}}{=} re^{i\varphi_r} \quad \in \mathbb{C}$
5. $S_{32} = S_{23} \stackrel{\text{def}}{=} te^{i\varphi_t} \quad \in \mathbb{C}$

With this first step done, we reduced the degrees of freedom, and we can write a mirror symmetric 3-port-device matrix as follows:

$$S_{3PD} = \begin{bmatrix} be^{i\varphi_b} & c_o e^{i\varphi_o} & c_o e^{i\varphi_o} \\ c_i e^{i\varphi_i} & re^{i\varphi_r} & te^{i\varphi_t} \\ c_i e^{i\varphi_i} & te^{i\varphi_t} & re^{i\varphi_r} \end{bmatrix} \qquad 31$$

3.1.2 Time reversal symmetry and energy conservation

We can group together those two symmetries since they are tightly related. In matrix form, those symmetries can be written as follow:

1. Time reversal: $S_{3PD}^{-1} = S_{3PD}$
2. Conservation of energy: $S_{3PD}^{H} \cdot S_{3PD} = \mathbb{1}$

Assuming true time-reversal symmetry, we are constraining $S_{3PD}$ to be Hermitian:

$$S_{3PD}^{H} = S_{3PD} \qquad 32$$

Therefore:

$$\begin{bmatrix} be^{-i\varphi_b} & c_i e^{-i\varphi_i} & c_i e^{-i\varphi_i} \\ c_o e^{-i\varphi_o} & re^{-i\varphi_r} & te^{-i\varphi_t} \\ c_o e^{-i\varphi_o} & te^{-i\varphi_t} & re^{-i\varphi_r} \end{bmatrix} = \begin{bmatrix} be^{i\varphi_b} & c_o e^{i\varphi_o} & c_o e^{i\varphi_o} \\ c_i e^{i\varphi_i} & re^{i\varphi_r} & te^{i\varphi_t} \\ c_i e^{i\varphi_i} & te^{i\varphi_t} & re^{i\varphi_r} \end{bmatrix}$$

Therefore, it follows:

1. $\varphi_b = 0$
2. $\varphi_r = 0$
3. $\varphi_t = 0$
4. $\varphi = \varphi_o = -\varphi_i$
5. $c = c_o = c_i$

In this way, we can write all the remaining degrees of freedom as positive real numbers: $b, r, t, c \in \mathbb{R}^+$ and a phase $\varphi \in [0, 2\pi]$.

At this stage, the Hermitian mirror symmetric scattering matrix can be written as follows:

$$S_{3PD} = \begin{bmatrix} b & ce^{i\varphi} & ce^{i\varphi} \\ ce^{-i\varphi} & r & t \\ ce^{-i\varphi} & t & r \end{bmatrix} \qquad 33$$

Before proceeding further, it is crucial to provide a comprehensive explanation of the significance of each degree of freedom. This is necessary to establish a clear physical understanding and address any ambiguities that may arise during subsequent discussions. By substituting $S_{3PD}$ of equation (33) into equation (28), we can see how the incident electric fields $E_{T,L,R}^i$ maps to the scattered one $E_{T,L,R}^s$ and we can deduce that:

1. $r$ and $t$ are the parameters that control the coupling between the left and the right part of the polaritonic channel through the 3-port-device. The parameter $t$ represents how the incident field incoming from the left (right) side of the 3-port-device is transmitted to the right (left) side. In the same way, we can understand that the parameter $r$ represents how the incident field incoming from the left (right) side is reflected.

2. The complex number $ce^{\pm i\varphi}$ represents how the far-field light couples to the near-field one that is propagating in the polaritonic channel and vice versa.
3. The parameter $b$ represents how the field impinging from the far-field region is reflected back.

Essentially, when the interaction between the scattering object and the polaritonic channel is minimal ($c \to 0$), the polaritonic channel should remain undisturbed. This implies that the 3-port device must exhibit unitary transmission and zero reflection ($t \to 1, r \to 0$).

We can continue the formulation by applying the time-reversal symmetry: $S_{3PD}{}^2 = \mathbb{1}$

$$\begin{bmatrix} b & ce^{i\varphi} & ce^{i\varphi} \\ ce^{-i\varphi} & r & t \\ ce^{-i\varphi} & t & r \end{bmatrix} \begin{bmatrix} b & ce^{i\varphi} & ce^{i\varphi} \\ ce^{-i\varphi} & r & t \\ ce^{-i\varphi} & t & r \end{bmatrix} = \begin{bmatrix} 1 & 0 & 0 \\ 0 & 1 & 0 \\ 0 & 0 & 1 \end{bmatrix} \qquad 34$$

From equation (34), we have a set of four unique conditions that our remaining degrees of freedom must fulfil:

1. $b^2 + 2c^2 = 1$
2. $c^2 + 2rt = 0$
3. $c^2 + r^2 + t^2 = 1$
4. $ce^{i\varphi}(b + r + t) = 0$

It is evident that we have four equations representing five degrees of freedom, resulting in one remaining free parameter. To facilitate the subsequent modelling of the near-field probe and maintain simplicity, we have opted to express everything in terms of the parameter $c$.

1. $b_\pm = \pm\sqrt{1 - 2c^2} \qquad c \in \left[-\frac{1}{\sqrt{2}}, \frac{1}{\sqrt{2}}\right] \wedge c \geq 0.$
2. $rt = -\frac{c^2}{2}$
3. $r^2 + t^2 = 1 - c^2$
4. $ce^{i\varphi}(b_\pm + r + t) = 0$

To solve this system of equations, we must consider two cases: $c = 0$ or $c \in (0, \frac{1}{\sqrt{2}}]$

**If $c = 0$**

1. $b_\pm = \pm 1$
2. $rt = 0$
3. $r^2 + t^2 = 1$

In this case we have to solve the ambiguity given by the equations in point 2 and 3. As discussed before, $r$ must be equal to 0, and $t$ must be equal to 1 in the limit of $c = 0$. However, the ambiguity in the parameter $b$ still remains unresolved:

$$r = 0, \quad t = 1, \quad b_\pm = \pm 1 \qquad 35$$

Therefore:

$$S_{3PD} = \begin{bmatrix} \pm 1 & 0 & 0 \\ 0 & 0 & 1 \\ 0 & 1 & 0 \end{bmatrix}$$

**If $c \neq 0 \wedge c \in \left[0, \frac{1}{\sqrt{2}}\right]$**

1. $b_\pm = \pm\sqrt{1 - 2c^2}$
2. $rt = -\frac{c^2}{2}$
3. $r^2 + t^2 = 1 - c^2$
4. $ce^{i\varphi}(b_\pm + r + t) = 0 \to b_\pm + r + t = 0 \to r = -(b_\pm + t)$

Let us focus on the equation of point 2 after the substitution of the equation of point 4:

$$-(b_\pm + t)t = -\frac{c^2}{2}$$

Solving this equation gives two solutions for the degree of freedom $t$:

$$-t^2 - b_\pm t + \frac{c^2}{2} = 0 \Rightarrow t_\pm = \frac{b_\pm \pm \sqrt{b_\pm^2 + 2c^2}}{-2} \Rightarrow t_\pm = \frac{b_\pm \pm 1}{-2}$$

Here we summarise our solutions:

1. $b_\pm = \pm\sqrt{1 - 2c^2}$
2. $t_\pm = \frac{b_\pm \pm 1}{-2}$
3. $r = -(b_\pm + t_\pm)$

We have to solve the ambiguity on $b_\pm$ and $t_\pm$. To do that, we must examine all the possible cases and consider the physical meaning of each situation when varying the parameter c within its existence boundaries $\left(0, \frac{1}{\sqrt{2}}\right]$.

As described earlier, in a physical situation, we must recover the fact that when the coupling coefficient $c \to 0$, the three-port device should not produce any effect in the propagation of the polaritons ($r \to 0$ and $t \to 1$). All the possible scenarios are shown in Fig. 2, and according to our physical criterion, the ambiguity is univocally solved by choosing $(b_-, t_-)$ (see Fig. 2D). For the sake of completeness, we can also solve the remaining ambiguity of $b_\pm$ in equation (35) by imposing the continuity of this parameter between the cases $c = 0$ and $c \neq 0 \wedge c \in \left[0, \frac{1}{\sqrt{2}}\right]$.

Finally, our 3-port-device scattering matrix can be expressed as a function of only two degrees of freedom $c$ and $\varphi$ that represent the amplitude and the phase of the coupling between the incident far-field light and the polaritonic modes.

$$S_{3PD}(c, \varphi) = \begin{bmatrix} b(c) & ce^{i\varphi} & ce^{i\varphi} \\ ce^{-i\varphi} & r(c) & t(c) \\ ce^{-i\varphi} & t(c) & r(c) \end{bmatrix} \quad\quad 36$$

Here:

$$b(c) = -\sqrt{1 - 2c^2} \quad\quad 37$$

$$t(c) = \frac{\sqrt{1 - 2c^2} + 1}{2} \quad\quad 38$$

$$r(c) = \sqrt{1 - 2c^2} - \frac{\sqrt{1 - 2c^2} + 1}{2} \quad\quad 39$$

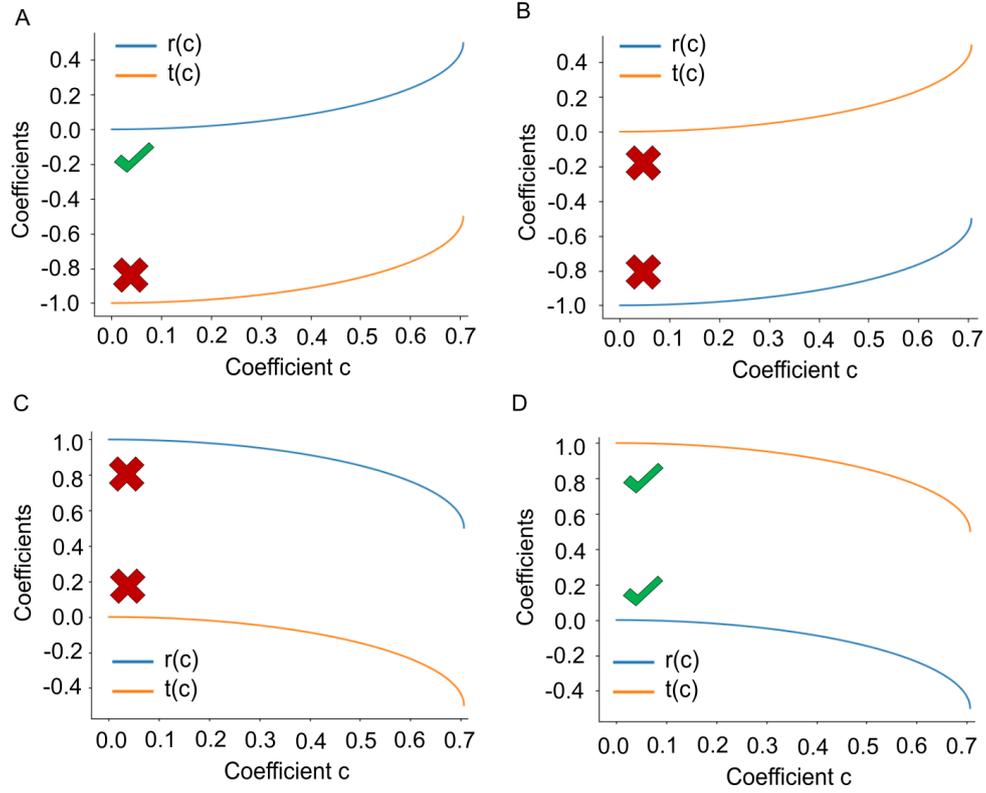

**Fig. 2, simulations of the 3x3 scattering matrix parameters $r$ and $t$ for $c \to 0$.** (**A**) Values of the function $r(c)$ and $t(c)$ are calculated in existence boundaries of $c$ when $b_+$ and $t_+$ are chosen. For $c \to 0$, $r \to 0$ converging to a physically meaningful value; however, $t \to -1$ suggests that with this sign choice, there is a $\pi$ phase jump when the polaritons propagate from one side to the other of the 1D channel, which is unphysical. (**B**) Same as panel A, but $b_+$ and $t_-$ are chosen. In this case, the sign selection suggests that in the absence of near-field to far-field coupling, the polaritons propagating from one side to the other of the 1D channel experience a total reflection, which is unphysical. (**C**) Same as panel A, but $b_-$ and $t_+$ are chosen. The sign selection leads to the same unphysical situation described in panel B. (**D**) Same as panel A, but $b_-$ and $t_-$ are preferred. With this sign selection, both $r$ and $t$ converge to a physically meaningful limit.

## 3.2 Modelling the s-SNOM near-field interaction

In order to complete our scattering theory of the near-field interaction in s-SNOM, we need to model the behaviour of the coupling coefficient $c$. This coefficient changes in function of the incident light wavelength and the tip-substrate distance. Considering that the s-SNOM tip can be approximated as a vertical electric dipole and polaritons exhibit exponentially decaying tails, we can model the coupling between the tip and polaritons using the following approach:

$$c(z) = C_0 e^{-\frac{z}{l}} \qquad 40$$

In equation (40), $C_0$ is the coupling coefficient amplitude, $z$ is the tip-substrate distance, and $l$ is a length scale that depends on the extension of the polariton tail above the substrate. To implement a realistic coupling, we must take into account that $l$ is a quantity that depends on the polariton wavelength $\lambda_P$. The longer the wavelength, the longer the polaritonic tails. In general, due to the exponential decay of the electric field outside the polaritonic waveguide, we can write:

$$l(\lambda_P) \propto \frac{\lambda_P}{2\pi} \qquad 41$$

On the other hand, the coupling coefficient $C_0$ depends on the electromagnetic response of the tip in function of the illumination wavelength. Such optical response can be written in the function of the incident vacuum light wavelength $\lambda_0$ and the tip apex radius $R$ as follow [32]:

$$C_0(\lambda_0) \propto (2\pi\lambda_0)^2 \cdot e^{-4\pi\lambda_0 R} \qquad 42$$

Finally, to evaluate the near-field observables, we need to simulate our system at many tip-substrate distances, mimicking the fact that the near-field probe undergoes subwavelength harmonic oscillations with frequency $\Omega$ above the substrate. Therefore, we need to add an explicit time dependence to the tip-substrate distance $z$ that reads:

$$z(t) = \Delta z \cdot \sin(2\pi\Omega t) + z_0 \qquad 43$$

At this stage, we can plug equation (43) in expression (40), completing the model of the scattering near-field probe coupled to a polaritonic system and, by evaluating at any given time $t$ the global scattering matrix of the 3-port-system (23), we can calculate the scattered electric field reaching the detector of the s-SNOM microscope:

$$E^s = S_{11}^G(t) \cdot E^i \qquad 44$$

## 3.3 Calculating the near-field observable

In s-SNOM measurements, the near-field scattered light sums up with a background electric field and is collected into a photodetector. The time-dependent signal intensity can be written as follows:

$$I(t) = \left| E^s(t) + E_{BG} e^{i\phi_{BG}} \right|^2 \qquad 45$$

From our model perspective, with this step, we introduced two new parameters ($E_{BG}, \phi_{BG}$) and a squared module operation. Notice that if $E_{BG}$ is set to 0, we are calculating a signal equivalent to the one measured with an ideal pseudo-heterodyne detection scheme [33] simply because this detection scheme suppresses the far-field background. On the other hand, if $E_{BG}$ is different from 0, we are calculating a signal equivalent to the one measured with the self-homodyne detection scheme. Therefore, by adding these two background parameters, the model is able to capture the near-field response in both pseudo-heterodyne and self-homodyne detection schemes.

Finally, in an s-SNOM apparatus, the electrical signal generated from the photodetector is demodulated at an integer harmonic of the tip oscillation frequency Ω with a lock-in amplifier. This means that, with our model, we can simulate the near-field observable $O_n$ by computing the n$^{th}$ Fourier coefficient of the simulated time-dependent signal $I(t)$.

## 4. Near-field simulations

To examine the validity of the proposed scattering theory, we present two experimental scenarios. The first one involves the simulation of polaritonic interference fringes, while the second focuses on the simulation of a 1D polaritonic lattice. The first scenario serves as a didactic demonstration, illustrating the limitations of the model and highlighting potential solutions. In contrast, the second scenario demonstrates the practical application of the model, addressing the main challenge that serves as the motivation for this research.

### 4.1 Polaritonic fringes

We first consider a polaritonic medium that exhibits a specific polaritonic dispersion $\lambda_P(\omega)$. In many experimental cases, this polaritonic medium is a finite piece of material that possesses physical edges that enable the observation of polaritonic fringes. To illustrate this scenario, Fig. 3A depicts the cross-section of a hexagonal boron nitride (hBN) flake with a thickness of 38nm that is positioned on top of a gold substrate.

Fig. 3B shows the s-SNOM measurement taken at the flake's natural edge. This measurement involves the simultaneous collection of both the spectral and spatial near-field responses in the vicinity of the edge. To accomplish this, the near-field probe scans the flake along a single line that is perpendicular to the flake's edge while the incident light wavelength is adjusted incrementally during the scanning process. Subsequently, we employ our scattering theory to simulate the near-field response, accounting for the fundamental polaritonic mode in hBN. As shown in Fig. 3C, the comparison between the experimental and simulated polaritonic fringes reveals a nice agreement. There is a consistent match between the experimental and simulated polaritonic fringes. However, the decay length of the simulated fringes does not align with the measured one.

This significant discrepancy stems from the limitations of our model, which only simulates perfect one-dimensional systems and do not accurately correspond to our experimental case. In the real case scenario, the measured fringes are influenced by the so-called geometrical losses caused by the fact that the polaritons are launched in a two-dimensional flake. Therefore, a genuine one-dimensional model fails to account for such losses. In this simple scenario, we can rectify this issue by directly considering the geometrical decay scales as $1/\sqrt[2]{x}$. Alternatively, for more complex systems, achieving a precise solution may be challenging.

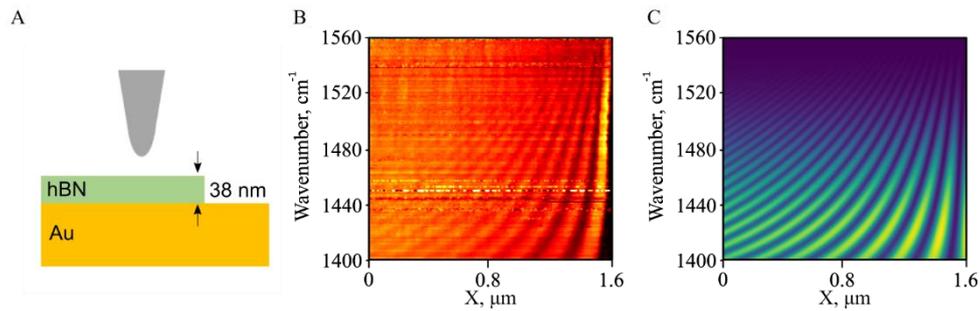

**Fig. 3, Experimental and simulated polaritonic fringes.** (**A**) Cross-section schematic of the hBN flake's edge. The flake (green) is placed over the gold substrate (yellow). In the schematic, the apex of the near-field probe is visible as the grey elongated shape floating in the proximity of the flake. (**B**) Pseudo heterodyne near-field scan of the polaritonic fringes. The polaritonic standing waves' wavelength becomes shorter as the illumination wavenumber increase (y-axis). (**C**) Near-field simulations of the experimental scenario of (B). The model is not affected by geometrical losses, and the polaritonic standing waves wavelength propagates longer.

## 4.2 Polaritonic 1D lattice

The focus of this second scenario is to simulate a one-dimensional polaritonic structure and illustrate how the underlying physics of the system can be mapped to the near-field response.

Let's consider a polaritonic waveguide composed of a periodic arrangement of alternating sections, as depicted in Fig. 4A. In this case, we can employ standard TMM approaches to calculate the properties of this 1D lattice. To study the physical properties of such an arrangement, we have the option to simulate a finite lattice and examine the reflection amplitude as a function of frequency. Alternatively, we can calculate the band structure of the infinite system by using TMM with periodic boundary conditions to the unit cell of the structure.

Fig. 4B shows the reflection response of the 1D structure, demonstrating the formation of band gaps. These band gaps are visible as regions for which the spectrum is characterised by unity reflection amplitudes. This is confirmed by comparing this result with the simulated band structure of the system, as depicted in Fig. 4C. Here, the polaritonic dispersion is folded within the First Brillouin zone, and the band-gaps observed in Fig. 4B match with the one shown in Fig. 4C.

Now that the underlying band structure is evaluated, we can now simulate the near-field response of the structure. This is shown in Fig. 4D, where we observe a clear correspondence between the near-field response and the underlying band structure presented in Fig. 4B-C. The regions without a near-field response correspond to the band gaps of the system, while the bright regions correspond to the polaritonic bands. This result not only establishes a connection between the underlying physics and the near-field response but also enables the extraction of other relevant information, such as the distribution of the electromagnetic field in different bands of the system.

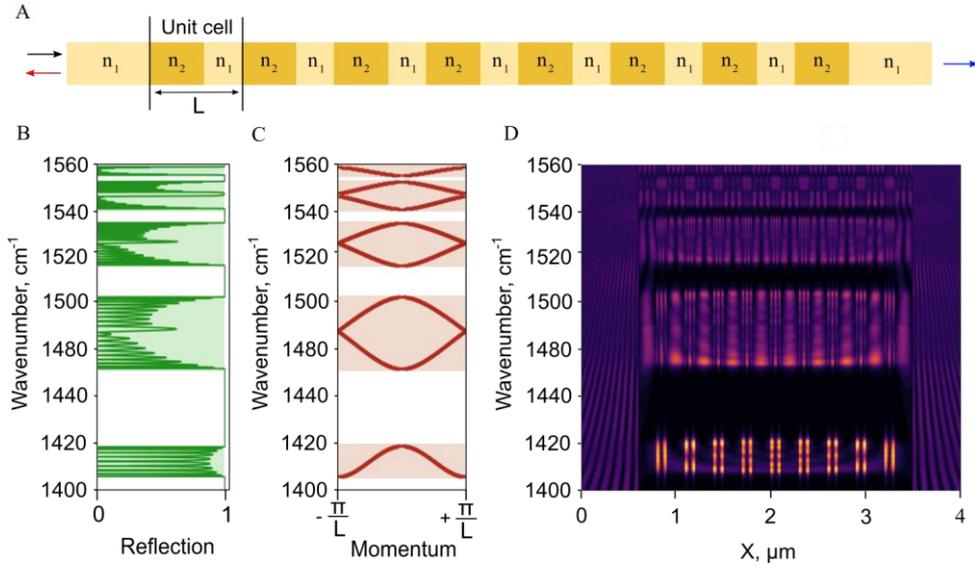

**Fig. 4, Simulation of a 1D polaritonic lattice.** (**A**) A schematic representation of the 1D lattice is shown, consisting of periodic repetitions of segments in the polaritonic structure characterised by different effective refractive indices, $n_1$ and $n_2$. The incident polaritonic amplitude (black arrow), reflected amplitude (red arrow), and transmitted amplitude (blue arrow) are depicted by arrows at the edges of the system. The unit cell of the system, with a length of $L$, is highlighted between two vertical black lines. (**B**) Simulated reflectivity using TMM for the finite lattice presented in (A) under lossless conditions. The green-shaded regions correspond to the bands of the system. Regions with reflectivity = 1 represent the polaritonic band gaps. (**C**) TMM simulation of the band structure for the unit cell presented in (A) with periodic boundary conditions under lossless conditions. (**D**) Near-field simulations of the scenario are presented in (A). In the centre of the scan, the near-field response of the lattice is visible, characterised by band gaps (black regions) and polaritonic bands (bright regions). The polaritonic bands exhibit strong frequency dependence, revealing the internal structure of the polaritonic modes. On the left and right sides of the scan, two regions without underlying periodicity are visible. Here, standard polaritonic fringes are observed, as the polaritonic lattice acts as a reflector when the tip scans the outer regions, allowing the formation and detection of standing waves.

## 5. Conclusions

In summary, we have developed a customised version of the transfer matrix method that effectively simulates the complex near-field response of 1D polaritonic systems. This method serves as a valuable tool in Nanophotonics, enhancing the understanding of information-rich near-field images in polaritonic systems. Furthermore, the insights gained from studying 1D systems can be qualitatively extended to more complex 2D systems.

To further refine this scattering theory, future research can focus on integrating an improved model for the coupling coefficient and polaritonic propagation. Additionally, this scattering theory can be expanded to include the response of non-polaritonic substrates by integrating other analytical or semi-analytical models, such as [34].

Looking beyond our specific findings, the versatility and intuitive nature of the TMM allow for the application of this scattering theory to various systems, including atoms or nanoparticles in close proximity to a waveguide. The case of propagating polaritons can be generalised to encompass generic guided modes, while the specific case of tip-apex near-field interaction can be extended to include generic scattering scenarios involving atoms or nanoparticles near a waveguide. By expanding the dimensionality of the scattering matrices, it becomes possible to explore different scenarios involving one or multiple scatterers interacting with TE and TM modes in a waveguide. These investigations can be conducted in both periodic and non-periodic arrangements of scatterers along the waveguide, ultimately facilitating the study of complex system responses and their correlation with relevant physical parameters.